\documentclass[useAMS,usenatbib]{mn2e}
\usepackage{aas_macros,graphicx,times,multirow,amsmath}
\title[A 6.4~h BH X-ray binary in NGC\,4490]{Discovery of a 6.4~h black hole binary in NGC\,4490}
\author[P.~Esposito et al.] {P.~Esposito,$^{1}$\thanks{E-mail: paoloesp@iasf-milano.inaf.it} G.~L.~Israel,$^{2}$ L.~Sidoli,$^1$ M.~Mapelli,$^{3}$ L.~Zampieri$^{3}$ and S.~E.~Motta$^4$\smallskip\\
$^1$Istituto di Astrofisica Spaziale e Fisica Cosmica - Milano, INAF, via E. Bassini 15, I-20133 Milano, Italy\\
$^2$Osservatorio Astronomico di Roma, INAF, via Frascati 33, I-00040 Monteporzio Catone, Italy\\
$^3$Osservatorio Astronomico di Padova, INAF, vicolo dell'Osservatorio 5, I-35122 Padova, Italy\\
$^4$European Space Astronomy Centre, ESA, PO Box 78, E-28691 Villanueva de la Ca\~{n}ada, Madrid, Spain
}
\date{Accepted 2013 September 24.  Received 2013 September 24; in original form 2013 September 5} \pagerange{\pageref{firstpage}--\pageref{lastpage}} \pubyear{2013}

\def\LaTeX{L\kern-.36em\raise.3ex\hbox{a}\kern-.15em
    T\kern-.1667em\lower.7ex\hbox{E}\kern-.125emX}

\def\xmm {\emph{XMM--Newton}}
\def\cxo {\emph{Chandra}}

\def\rst {\emph{ROSAT}}

\def\src {CXOU\,J123030.3+413853}
\def\flux {\mbox{erg cm$^{-2}$ s$^{-1}$}}
\def\lum {\mbox{erg s$^{-1}$}}
\def\nh {$N_{\rm H}$}

\begin{document}

\label{firstpage}
\maketitle
\begin{abstract}
We report on the discovery with \cxo\ of a strong modulation ($\sim$90\% pulsed fraction) at $\sim$6.4~h from the source \src\ in the star-forming, low-metallicity spiral galaxy NGC\,4490, which is interacting with the irregular companion NGC\,4485. This modulation, confirmed also by \xmm\ observations, is interpreted as the orbital period of a binary system. The spectra from the \cxo\ and \xmm\ observations can be described by a power-law model with photon index $\Gamma\sim1.5$. During these observations, which span from 2000 November to 2008 May, the source showed a long-term luminosity variability by a factor of $\sim$5, between $\sim$$2\times10^{38}$ and $1.1\times10^{39}$~\lum\  (for a distance of 8 Mpc). The maximum X-ray luminosity, exceeding by far the Eddington limit of a neutron star, indicates that the accretor is a black hole. Given the high X-ray luminosity, the short orbital period and the morphology of the orbital light curve, we favour an interpretation of \src\ as a rare high-mass X-ray binary system with a Wolf--Rayet star as a donor, similar to Cyg\,X--3. This would be the fourth system of this kind known in the local Universe. \src\ can also be considered as a transitional object between high mass X-ray binaries and ultraluminous X-ray sources (ULXs), the study of which may reveal how the properties of persistent black-hole binaries evolve entering the ULX regime.
\end{abstract}

\section{Introduction}

NGC\,4490, at a distance of 7--10 Mpc,\footnote{Distances from the NASA/IPAC Extragalactic Database (NED), see \mbox{http://ned.ipac.caltech.edu/}.} is a spiral galaxy interacting with the irregular galaxy NGC\,4485 (\citealt*{devaucouleurs76}; \citealt{tully88,devaucouleurs91}). At 8 Mpc (the distance we assume throughout, following many previous studies) the angular separation between the pair ($\sim$3.6 arcmin) corresponds to a projected distance of $\sim$8 kpc, while the sizes of NGC\,4490 and NGC\,4485 are $\sim$15 and $\sim$5.6 kpc, respectively \citep*{clemens99}. Radio and infrared observations indicate an enhanced star-forming activity (\citealt*{viallefond80}; \citealt{duval81,klein83,thronson89}) at a rate of $\approx$5~M$_{\sun}$~yr$^{-1}$ \citep{clemens02}.

The first X-ray observations of NGC\,4485/4490 with good angular resolution were carried out with \rst\ and resulted in the detection of 5 compact X-ray sources \citep*{read97,roberts00,liu05}. Subsequent observations with \cxo\ increased the number of resolved sources to 38 \citep{roberts02,fridriksson08,richings10}. Among these sources there is \src, which was discovered in NGC\,4490 by \citet{roberts02} with the first \cxo\ observation, performed in 2000 November. The object is located at $\sim$1.1 arcmin ($\sim$2.5~kpc) from the optical nucleus of NGC\,4490. In a study of the variability of the X-ray sources of NGC\,4490/4485, \citet{fridriksson08} observed both short- and long-term flux variability in \src\ but, overall, since its discovery the source did not attract particular interest. 

Here we report on a comprehensive analysis of all the \cxo\ and \xmm\ data of \src. We present evidence that its X-ray emission is modulated at a period of $\sim$6.4 hours ($\sim$23~ks) which can be interpreted as the orbital period of a binary system. We also found that in several observations \src\ approached, and perhaps exceeded, an X-ray luminosity ($L_{\mathrm{X}}$) of $\sim$$10^{39}$~\lum, which is the traditional threshold for ultraluminous X-ray sources (ULXs, see \citealt{feng11} for a recent review). These sources are matter of debate as they may harbour intermediate-mass ($\approx$$10^2$--$10^5$~M$_{\sun}$) black holes (BHs), bridging the gap between stellar-mass BHs and the super-massive BHs found in galactic nuclei (see e.g. \citealt{miller04,vandermarel04}). However, apart from a handful of extremely luminous ($L_{\mathrm{X}}>5\times 10^{40}$~\lum; \citealt{sutton12}) or hyperluminous ($L_{\mathrm{X}}>10^{41}$~\lum; \citealt{farrell09}) objects, observationally BHs of several hundreds to thousands M$_{\sun}$ 
are not required for the majority of ULXs (e.g. \citealt{fabbiano05}), which remain consistent with alternative interpretations -- either unusually massive stellar BHs ($\sim$20--100~M$_{\sun}$) or a different accretion state (e.g. \citealt{zampieri09,feng11}).
\begin{figure*}
\centering
\resizebox{\hsize}{!}{\includegraphics[angle=0]{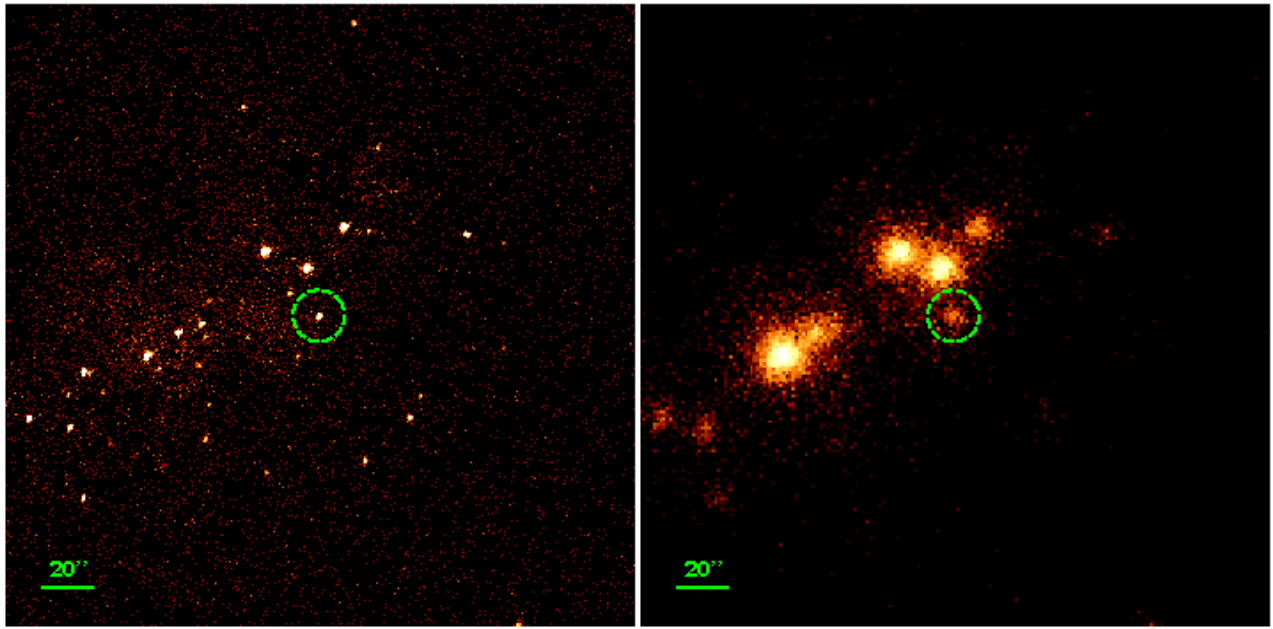}}
\caption{\label{ds9} Field of \src, marked by the circle (20~arcsec diameter) as imaged by \cxo\ (left) and \xmm\ (right) in the 0.3--8~keV energy band (all observations were combined). Each image side is 4~arcmin wide. X-ray emission from the hot interstellar medium of NGC\,4485/4490 is apparent in both images.}
\end{figure*}

\section{Observations}

\cxo\ observed the galaxy pair NGC\,4485/4490 with its Advanced CCD Imaging Spectrometer (ACIS; \citealt{garmire03}) instrument three times (between 2000 and 2004), for a total exposure of approximately 100 ks \citep{roberts02,fridriksson08,gladstone09,richings10,yoshida10}. Details of the observations are summarized in Table\,\ref{logs}. In all observations the ACIS was operated in the standard timed exposure full-frame mode, with a CCD readout time of 3.24~s. In the first observation the events were telemetered in `faint' mode, while the other two observations were performed with the `very faint' telemetry format. Each time, \src\ was positioned on the back-illuminated chip S3. The data were reprocessed with the \cxo\ Interactive Analysis of Observations software  package (\textsc{ciao}, version 4.5). Fig.\,\ref{ds9} shows the field of \src\ in the 0.3--8 keV energy band. The source counts were selected in an $\sim$2-arcsec radius region and the background in an annulus with inner (outer) radius of 5 (10) arcsec. The spectra, the ancillary response file and the spectral redistribution matrices were created using \textsc{specextract}.

\xmm\ observed  NGC\,4485/4490 two times: in 2002, for about 18~ks \citep{winter06,gladstone09,yoshida10}, and in 2008, for 37~ks (Table\,\ref{logs}). For this work we used only the data from the European Photon Imaging Camera, which consist of two MOS \citep{turner01short} and one pn \citep{struder01short} CCD cameras. The front-illuminated MOS cameras were operated both times in full frame mode (integration time: 2.6~s), while the back-illuminated pn was set in `extended' full frame mode in 2002 (integration time: 0.2~s) and in full frame in 2008 (integration time: 73~ms); the medium optical blocking filters were used for all observations. Both observations (but particularly the second) were affected by rather intense soft-proton flares. Since the removal of the intervals of flaring background significantly alters the light curves, we chose to clean the data for the spectral analysis (by applying intensity filters to the event lists) and to limit the timing analysis to the unfiltered MOS data, since the MOS cameras are less affected by proton flares. In order to reduce the contamination from the hot interstellar medium emission of NGC\,4490 and from nearby sources (Fig.\,\ref{ds9}), the source photons were accumulated for each camera in small circular regions (10-arcsec radius, $\sim$60\% of the enclosed energy fraction at 1.5 keV). The background counts were estimated from source-free composite regions in the same chip as the source. The data were processed using the \xmm\ Science Analysis Software (\textsc{sas}, version 12). The ancillary response files and the spectral redistribution matrices for the spectral analysis were generated with \textsc{arfgen} and \textsc{rmfgen}, respectively.
\begin{table}
\centering \caption{Summary of the observations used in this work.} \label{logs}
\begin{tabular}{@{}lccc}
\hline
Mission~/~Obs.\,ID  & Date & Exp. & Count rate$^{a}$\\
 & & (ks) & (count~s$^{-1}$) \\
\hline
\cxo~/ 1579 & 2000 Nov 03 & 19.5 & $(8.2\pm0.7)\times10^{-3}$ \\
\emph{XMM}~/~0112280201 & 2002 May 27 & 17.5 & $(1.1\pm0.1)\times10^{-2}$ \\
\cxo~/~4725 & 2004 Jul 29--30 & 38.5 &  $(9.4\pm0.5)\times10^{-3}$ \\
\cxo~/~4726 & 2004 Nov 20 & 39.6& $(1.6\pm0.2)\times10^{-3}$ \\
\emph{XMM}~/~0556300101 & 2008 May 19 & 37.4 &$(1.1\pm0.1)\times10^{-2}$ \\
\hline
\end{tabular}
\begin{list}{}{}
\item[$^{a}$] Net source count rate in the 0.3--8~keV energy band using the extraction regions described in the text; for the \xmm\ observations we give the pn rate. The values are not corrected for point-spread function and vignetting effects.
\end{list}
\end{table}

\section{Analysis and results}

\subsection{Discovery of the modulation and timing analysis}

\src\ is one of the about 30 new X-ray pulsators discovered so far by the \cxo\ Timing Survey at Brera And Roma astronomical observatories project (CATS@BAR; Israel et al., in preparation; see also \citealt{eis13,eisrc13}). CATS@BAR is aimed at the exploitation of the ACIS archival data (timed exposure imaging observations, with no gratings in use). As of 2013 May 31, approximately 8\,900 observations were retrieved and about 415\,000 light curves were extracted. For the $\sim$87\,000 light curves with more than $\sim$150 photons, Fourier  power spectra were computed and searched for coherent or quasi-coherent signals in an automatized way by applying a detection algorithm based on that described in \citet{israel96}.

\begin{figure}
\centering
\resizebox{\hsize}{!}{\includegraphics[angle=-90]{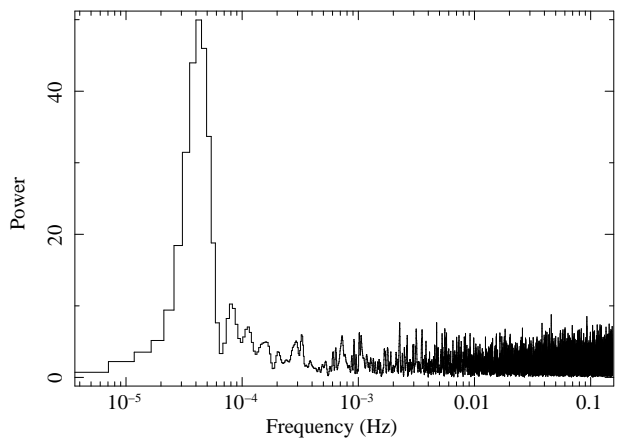}}
\caption{\label{powspec} Average power spectrum form the three \cxo\ observations. The $\sim$6.5~h ($\sim$$4.26\times10^{-5}$~Hz) signal is apparent.}
\end{figure}

In the case of \src, a promising signal was first found in the power spectrum of the observation 4725. It showed a single high peak at the second/third Fourier frequency (depending on the assumed number of bins in the time series), corresponding to a period of $\approx$6.5~h.  Considering the 8192 independent frequencies in the spectrum with frequency resolution of $\sim$$1.9\times10^{-5}$~Hz, the detection is significant at $\sim$6.6$\sigma$. When taking into account also the number of light curves searched in the CATS@BAR programme ($\sim$87\,000), the significance of the peak is  4.7$\sigma$. In order to better investigate the signal, we used the three \cxo\ observations (Table\,\ref{logs}) to produce a new power spectrum (Fig.\,\ref{powspec}), average of three power spectra (the power was normalized following \citealt{leahy83} and dividing the expectation value and standard deviation of the noise distribution by the number of averaged power spectra). The peak significance increased, as expected if the signal was present in all data sets, and was estimated to be $\sim$9$\sigma$. 

We carefully investigated the possibility of a signal of instrumental origin. A check for spurious signals that might be introduced in the \cxo\ light curves by the spacecraft dithering, is automatically performed by the CATS@BAR signal detection pipeline. The procedure is based on the \textsc{ciao} task \textsc{dither\_region}\footnote{See http://cxc.harvard.edu/ciao/ahelp/dither\_region.html.} and checks whether an artificial signal due to the variation of the fractional area for the given source extraction region is present in the original time series. The analysis gave a negative result: no spurious signal is present at or nearby the detected peak. As a further check, the nearby sources were searched for a similar modulation, with negative results.  Finally, we notice that the CATS@BAR pipeline preserves the information (frequencies, amplitudes, etc.) related to all the detected peaks, regardless of whether they are spurious or not. As of 2013 May 31 the search produced more than 155\,000 detected peaks from $\sim$87\,000 light curves. None of the spurious signals falls at the or close to the frequency of the peak of \src. We conclude that the signal (which is confirmed also by the 2008 \xmm\ observation, see below) is real and intrinsic  to \src. 

\begin{figure*}
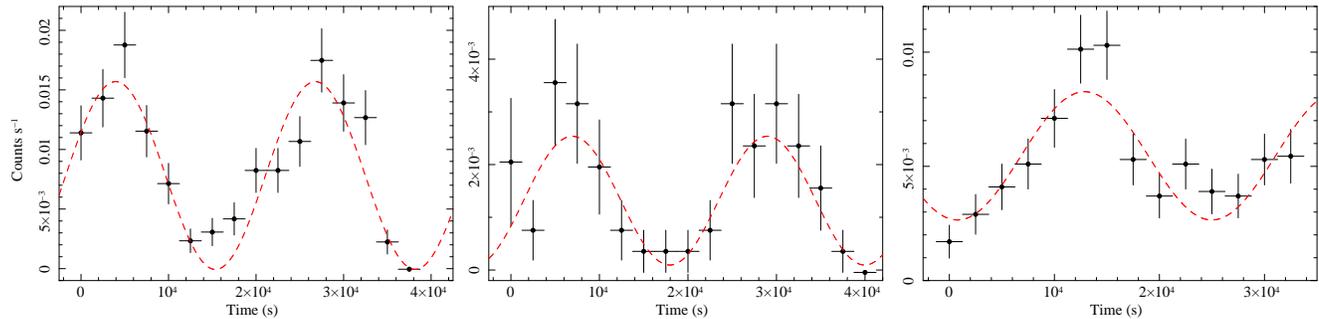

\centering
\resizebox{\hsize}{!}{\includegraphics[angle=-90]{lc_4725.eps}\hspace{-1cm}\includegraphics[angle=-90]{lc_4726.eps}\hspace{-1cm}\includegraphics[angle=-90]{lc_xmm2.eps}}
\caption{\label{profiles} Background-subtracted 0.3--8 keV light curves of \src\ from the \cxo\ 2004 July (left) and 2004 November (middle) observations and from the \xmm\ 2008 May observation (right). The red dashed line shows the best-fitting sinusoidal function for each data set.}
\end{figure*}

We then inspected the \cxo\ and \xmm\ light curves. As it can be seen from Fig.\,\ref{profiles}, the $\sim$6.5~h flux modulation is rather large and clearly recognizable in the three $\sim$40-ks observations (consistent flux variations are present also in the other pointings, although the short exposure times prevent an unambiguous association to the $\sim$6.5~h period). In each of these observations we measured the period by fitting a sinusoidal function to the light curve. The individual periods, ($6.32\pm0.14$)~h for the \cxo/4725 observation, ($6.11\pm0.36$)~h for the \cxo/4726, and ($6.73\pm0.38$)~h for the \xmm/0556300101 (MOS data), are all consistent within the uncertainties (all errors are 1$\sigma$).

\begin{figure}
\centering
\resizebox{\hsize}{!}{\includegraphics[angle=-90]{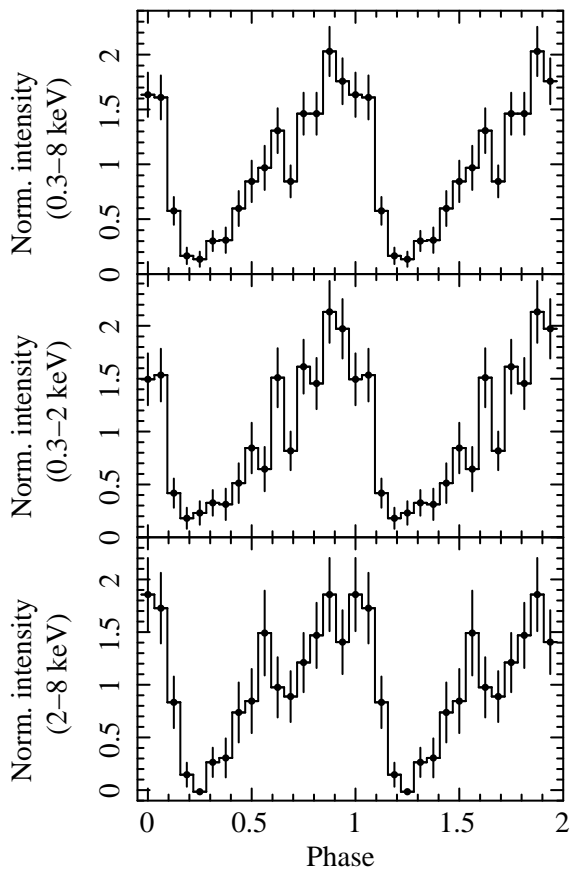}}
\caption{\label{cxofold} Background-subtracted folded profiles (from all \cxo\ observations combined) of \src\  in the total (0.3--8 keV, top), soft (0.3--2 keV, middle), and hard (2--8 keV) energy bands.}
\end{figure}

Folding the three \cxo\ observations with a common period $P$ and looking for the value that maximizes the pulsed fraction, we obtain $P=(6.43 \pm 0.11)$~h. In Fig.\,\ref{cxofold} we show the corresponding profile obtained by folding all the \cxo\ data. The profile is slightly asymmetric, and a double-sinusoidal function provides a somewhat better fit to it than a single one ($\chi^2_\nu=0.91$ for 11 degrees of freedom (dof) against 2.68 for 13 dof). The pulsed fraction, defined as $(M-m)/(M+m)$, where $M$ is the maximum of the pulse profile and $m$ the minimum, is ($88\pm6$)\% in the 0.3--8~keV band. We also show an example of soft (0.3--2~keV) and hard (2--8~keV) profiles. The pulsed fractions are compatible ($84\pm8$\% in the soft band and $85\pm11$\% in the hard) and the hardness ratio between the two profiles does not significantly deviate from a constant. More in general, no statistically significant variations of the profile are observed as a function of the energy in the \cxo\ data.

\subsection{Spectral analysis}

For the spectral analysis (performed with the \textsc{xspec} version 12.7 fitting package; \citealt{arnaud96}) we first concentrate on the ACIS data because, owing to the high angular resolution of the \cxo\ mirrors ($\sim$0.4 arcsec half-maximum) and the stable background during the exposures, they are those with the highest signal-to-noise ratio  ($\sim$97--99 per cent of source counts in each observation). Given the paucity of counts, especially in the observations 1579 and 4726, we fit simple models to the three spectra simultaneously, with the normalizations free to vary and the other parameters tied up between the data sets. A power law with photon index $\Gamma\sim1.5$ or a multi-colour disc \citep{mitsuda84short,makishima00short} with characteristic temperature $kT\sim2$~keV, both modified for the interstellar absorption, provide equally good fits to the data. The best-fitting parameters are given in Table\,\ref{specs}. The \nh\ value is only poorly constrained, but for both models the best-fitting value is substantially larger than the Galactic one ($1.8\times10^{20}$~cm$^{-2}$; \citealt{dickey90,kalberla05}) and comparable with that of the point sources in NGC\,4490 \citep{roberts02,gladstone09}.
\begin{table*}
\begin{minipage}{16cm}
\centering \caption{\cxo\ and \xmm\ spectral results. Errors are at a 1$\sigma$ confidence level for a single parameter of interest.} \label{specs}
\begin{tabular}{@{}lccccccc}
\hline
Model$^a$ & Obs.\,ID & \nh$^b$ & $\Gamma$ & $kT$ & Observed flux$^c$ & Luminosity$^d$ & $\chi^2_\nu$ (dof) \\
 & & ($10^{21}$ cm$^{-2}$) &  & (keV) &($10^{-14}$ \flux) & ($10^{38}$ \lum) &  \\
\hline
\multicolumn{8}{c}{\cxo}\\
\hline
 & 1579 & & & & $8.4^{+0.9}_{-1.0}$ & $9.3\pm0.9$\\
\textsc{phabs*powerlaw} & 4725 &  $3.5\pm1.5$ & $1.5\pm0.2$ &-- & $10.0\pm0.9$ & $11.1\pm0.8$ & 0.89 (20) \\
 & 4726 &  & & & $1.8^{+0.4}_{-0.3}$ &  $2.0^{+0.4}_{-0.3}$ \\
\hline
 & 1579 &  & & & $7.8\pm1.0$ & $6.8^{+1.0}_{-0.6}$\\
\textsc{phabs*diskbb} & 4725 &  $1.1^{+1.1}_{-1.0}$ & -- &  $1.9^{+0.4}_{-0.3}$ & $9.0^{+1.0}_{-0.8}$ & $7.9^{+0.9}_{-0.7}$ & 0.86 (20) \\
 & 4726 & & & & $1.6^{+0.4}_{-0.3}$ &  $1.4^{+0.3}_{-0.2}$  \\
\hline
\multicolumn{8}{c}{\xmm}\\
\hline
\multirow{2}{*}{\textsc{phabs*powerlaw}} & 0112280201 &  \multirow{2}{*}{$4.5^{+1.8}_{-1.5}$} & \multirow{2}{*}{$1.8^{+0.2}_{-0.3}$} & \multirow{2}{*}{--} & $7.1^{+1.2}_{-1.0}$ & $8.9^{+1.1}_{-1.0}$ & \multirow{2}{*}{0.81 (16)} \\
 & 0556300101 &  & & & $7.7\pm0.8$ &  $9.6^{+1.2}_{-1.0}$ \\
\hline
\multirow{2}{*}{\textsc{phabs*diskbb}} & 0112280201 &  \multirow{2}{*}{$1.8^{+1.2}_{-1.0}$} & \multirow{2}{*}{--} &  \multirow{2}{*}{$1.4^{+0.3}_{-0.2}$} & $6.4^{+1.0}_{-0.9}$ & $5.7^{+0.9}_{-0.7}$ & \multirow{2}{*}{0.80 (16)} \\
 & 0556300101 & & & & $7.0\pm0.8$ &  $6.3\pm0.6$  \\
\hline
\end{tabular}
\begin{list}{}{}
\item[$^{a}$] \textsc{xspec} model.
\item[$^{b}$]  The abundances used are those of \citet*{wilms00}. Essentially identical values are obtained with the abundances by \citet{lodders03} and \citet{asplund09}, while values $\sim$30\% lower are derived with those of \citet{anders89}. The photoelectric absorption cross-sections are from \citet{balucinska92}.
\item[$^{c}$] In the 0.3--8 keV energy range. 
\item[$^{d}$] Isotropic 0.3--10 keV luminosity, calculated from the unabsorbed flux assuming a distance of 8 Mpc.
\end{list}
\end{minipage}
\end{table*}

While the fluxes measured in the first two observations (2000 November and 2004 July) are compatible, the third \cxo\ observation (2004 November) caught \src\ at a significantly lower flux, implying a variability by a factor of $\sim$5. For a distance of 8~Mpc, the highest observed flux (obs. 4725) translates into a 0.3--10 keV isotropic luminosity of $\sim$$1.1\times10^{39}$~\lum\ for the power-law fit, and of  $\sim$$8\times10^{38}$~\lum\ for the multi-colour disc. We note that during that observation the flux at the modulation maximum was more than two times the average one, indicating an isotropic luminosity above $10^{39}$~\lum\ (in part of the orbit) also for the multicolour disc fit.

We performed a similar analysis with the \xmm\ data. After the proton flares cleaning, the net exposures were reduced in the first (second) observation to 11.2~ks (14.7~ks), 16.8~ks (22.0~ks), and 16.6~ks (23.7~ks) in the pn, MOS\,1, and MOS\,2 detectors, in the order given. The spectral parameters derived from the fits are consistent with those obtained from the \cxo\ observations. In both pointings the fluxes are  close to the values measured in the first two \cxo\ data sets, when the luminosity of \src\ was about $10^{39}$~\lum.

\section{Discussion}

During a systematic search for new X-ray pulsators in the \cxo\ ACIS archive, we discovered a strong ($\sim$90\% pulsed fraction) modulation at about 6.4~h from the source \src\ in NGC\,4490 (a star-forming galaxy, interacting with the companion NGC\,4485). This modulation is naturally interpreted as the orbital period of a binary system. The spectral analysis of the available data (three \cxo\ and two \xmm\ observations between 2000 and 2008; Table\,\ref{logs}) showed that \src\ is a rather bright and moderately-variable (within a factor of $\sim$5) source, with a maximum observed luminosity exceeding $10^{39}$~\lum.

The Eddington limit for spherical accretion of fully ionized hydrogen on to a compact object of mass $M$ is $L_{\mathrm{E}}\simeq1.3\times10^{38} M/\mathrm{M}_{\sun}$ \lum. If \src\ is Eddington limited, its mass should then be greater than $\sim$10~M$_{\sun}$ (or than $\sim$5~M$_{\sun}$ for a He or C/O donor, which is likely the case, see below), strongly arguing against a neutron star primary. 
In fact, it is not possible to completely rule out a neutron star on energetic grounds alone [the Eddington limit can be exceeded by a factor of a few even without invoking unusual accretion regimes, see e.g. \citealt*{grimm03}, and neutron-star binaries have been observed to achieve, albeit only briefly, luminosities approaching $\approx$$10^{39}$~\lum\  (during giant outbursts in Be X-ray binary systems, see e.g. \citealt{white78,reig07})], but persistent sources with luminosity above a few $10^{38}$~\lum\ are clearly much easier to explain in terms of BH binaries.
Moreover, the lack of breaks between $\sim$$5\times 10^{38}$ and $\ga$$2\times 10^{40}$~\lum\ in the X-ray luminosity functions (XLFs) for point sources observed in nearby spiral galaxies
suggest that above $\sim$$5\times10^{38}$~\lum\ XLFs are populated mainly by X-ray binaries with BH accretors (e.g. \citealt*{sarazin00,gilfanov04,kim04,fabbiano06,ivanova06}). We thus believe that a BH is indeed the most probable primary for \src, and in the following we will discuss the source in this context.

\subsection{The nature of the binary system}\label{binarysystem}

The X-ray periodicity of 6.4~h indicates an orbital modulation and, considering the high X-ray luminosity, there are two possibilities for the source nature: either a low mass (LMXB) or a high mass X-ray binary (HMXB). While the 0.3--10 keV spectral distribution is compatible with both kinds of X-ray binaries, the timing analysis (orbital period and morphology of the orbital light curve) discloses more information on the source nature.

In the first hypothesis, an orbital period of 6.4~h is typical of an LMXB with a late-spectral-type main sequence companion \citep{verbunt93}. Eclipsing and/or dipping LMXBs with similar orbital periods (e.g., MXB\,1659--298, with P$_{\mathrm{orb}}=7.11$~h; \citealt{cominsky84}) display X-ray emission  periodically modulated by the presence of X-ray eclipses produced by the low mass companion in a high inclination system ($i>70$\degr; e.g. \citealt*{frank87}), and/or by the presence of dips due to the obscuration produced by the accreting matter located in the disc bulge (see \citealt{diaztrigo06} for a review). However, in the so-called `dippers', the X-ray minima are typically much sharper than what shown by \src\ and, in any case, the morphology of the source light curve is very different.
Moreover,  dippers usually show energy-dependent X-ray emission, since the periodic modulation is produced by absorption of X-rays by the (neutral and ionized) matter  into the line of sight, contrary to what we observed in \src.

A smoother modulation in the X-ray light curve of an LMXB can be produced in a so-called accretion disc corona (ADC) source: the X-ray emission in this kind of X-ray binaries viewed edge-on is mostly blocked by the dense, obscuring, outer edge of the accretion disc. The presence of an extended corona scatters into the line of sight the X-ray emission coming from the central obscured source, so that the outer edge of the dense disc modulates this scattered X-ray emission, producing an observed orbital X-ray light curve with a very smooth profile \citep{mason82}. The prototype of the ADC sources is X\,1822--371, a neutron star LMXB with an orbital period of 5.57~h (e.g. \citealt{somero12} and references therein).
However, the profile morphology in ADC sources, although much smoother than in other eclipsing or dipping LMXBs,  is again very different from that of \src. To summarize, the hypothesis that \src\ is an LMXB seems to be unlikely, also because LMXBs containing BHs are usually transient X-ray sources, with outburst durations of the order of weeks to months (e.g. \citealt{remillard06}), whereas \src\ is persistent, although variable within a factor of 5 (see Table\,\ref{specs}). 

A viable alternative  is that \src\ is an HMXB. Given its short orbital period, both a main-sequence and a supergiant early-type companion cannot fit into the very narrow orbit. The remaining possibility is that the source is a late evolutionary product of an HMXB,  hosting a Wolf--Rayet (WR) star (see \citealt{crowther07} for a review on WR stars), similar to Cyg\,X--3 \citep{giacconi67}. 

Cyg\,X--3 is a bright X-ray binary ($L_{\mathrm{X}}\approx10^{38}$~\lum) that shows an even shorter orbital period of 4.8~h \citep{canizares73} and a remarkably similar orbital asymmetric X-ray light curve, with a slow rise and a fast decline \citep{zdziarski12}. Cyg\,X--3 is a unique source in our Galaxy, being the only X-ray binary containing a WR star \citep{vankerkwijk96}. The nature of the compact object accreting from the WR wind is still unclear, although there are many indications (radio, infrared and X-ray emission properties; \citealt*{szostek08,szm08}) pointing to a 2--5~M$_{\sun}$ BH (see, e.g., \citealt*{zdziarski13}), as also suggested by evolutionary models \citep{lommen05,belczynski13}. The peculiar shape of the Cyg\,X--3 orbital X-ray light curve and its remarkable energy independence received attention since early 1970s (e.g. \citealt*{hertz78} and references therein), and they were explained by  transfer of X-rays through a `cocoon', a cloud of scattering medium. A more recent study of the orbital modulation of X-rays in Cyg\,X--3 was performed by \citet{zdziarski10}, finding that the X-ray modulation can be produced by both absorption and Compton scattering in the WR stellar wind (the minima being accounted for by the highest optical depth at the superior conjunction). In \src\ the shape of the orbital profile indicates an asymmetry in the scattering matter. This anisotropy is likely due to either a WR wind focused towards the compact object or a cloud of ionized matter associated with a bulge in the accretion disc, probably produced by the gravitationally focused stellar wind colliding with the outer accretion disc. Additional contribution by a jet from the putative BH is a possibility. Eclipses by the companion star can be smoothed out by Compton scattering with high optical depth (see, e.g., \citealt{hertz78}). The effect of photoelectric absorption of soft X-rays in the WR wind can be reduced by the possible presence of a soft excess due to re-emission of the absorbed continuum by the WR stellar wind. So, several contributions are possible which could modulate the X-ray emission along the orbit, similar to the luminous Galactic source \mbox{Cyg\,X--3.}

WR--BH binary systems are very rare. Apart from \mbox{Cyg\,X--3}, for which the nature of the compact object is still unclear, only two such systems are known  to date, both in external galaxies. One is located in the dwarf irregular galaxy IC\,10 (IC\,10~X--1; \citealt{bauer04,prestwich07}) and the other in the spiral NGC~300 (NGC\,300~X--1; \citealt{carpano07,cpp07}). In these two sources the orbital periods are much larger than in \src, exceeding 30~h \citep{prestwich07,crowther10}.\footnote{It is interesting to note that IC\,10~X--1 and NGC\,300~X--1 are among the most massive stellar-mass BHs for which dynamical mass constraints are available \citep{ozel10}. The most probable mass of NGC\,300~X--1 is $\approx$20~M$_{\sun}$ \citep{crowther10}, while that of IC\,10~X--1 is $\approx$33~M$_{\sun}$ \citep{prestwich07,silverman08}.} So, \src\ could be a candidate extragalactic twin to Cyg\,X--3, as well as one of the very few WR--BH binary systems known in the local Universe. The interest in the systems of this kind, besides their rarity, is also due to the fact that they might be the progenitors of a double-BH binary (\citealt{lommen05}; \citealt*{bulik11}). 

\subsection{A transitional object between BH X-ray binaries and ULXs}

As mentioned above, in at least one of the {\it Chandra} observations the luminosity of \src\ marginally crosses the conventional luminosity threshold of $10^{39}$~\lum\ and may then be defined as a `borderline' ULX. Other variable (and persistent) X-ray sources in nearby galaxies show large excursions in luminosity and enter only at times the low-luminosity ULX regime (e.g. NGC\,253~X--1; Pintore et al., submitted). In addition, a few transient ULXs, with luminosities at maximum slightly above $10^{39}$~\lum, have also been identified (e.g. \citealt{sivakoff08short,middleton12,middleton13short,soria12,esposito13}). At low luminosities some of these sources show properties consistent with those of more conventional X-ray binaries. Interestingly, at least one of the transient ULXs in M\,31 was identified with an LMXB \citep{middleton12}, while \src\ is likely accreting from a WR star (see Section\,\ref{binarysystem}), consistent with the persistent character of the source. The existence of these transitional objects is clearly very interesting because they reveal how the properties of BH binaries evolve entering the ULX regime. \src\ is particularly interesting in this respect because it shows evidence of an orbital periodicity, detected only in a handful of ULXs (\citealt*{kaaret07,liu09}; \citealt{zampieri12}), which may make it possible to extract additional physical information about the system.

The observed X-ray spectra of \src\ do not have sufficient counting statistics to discriminate among different models and are broadly consistent with several interpretations. The power-law fit gives spectral indices reminiscent of the hard state of Galactic BH binaries, but at the same time the \textsc{diskbb} fits give temperatures consistent with those of the soft state. In general, the spectra appear slightly harder than those of other low luminosity ULXs (e.g. \citealt{stobbart06,berghea08}). However, higher counting statistics observations are needed to better constrain the spectral properties and understand if the source may show spectral signatures and/or transitions similar to those typically observed in brighter ULXs \citep{grd09,kajava09,pintore12}. Assuming that at peak the source is in a soft/disc dominant state and using the normalization $K$ of the \cxo\ \textsc{diskbb} fit, it is possible to estimate the BH mass as \citep{lorenzin09}: 
$$M\approx 12.5\bigg(\frac{D}{1~\mathrm{Mpc}}\bigg)\bigg(\frac{K}{\cos i}\bigg)^{1/2}\mathrm{M}_{\sun},$$ where $i$ is the inclination angle ($i=0$ corresponds to a face-on disc). For \src\ (obs.~4725), one gets $M\approx2.8\,(D_8/\sqrt{\cos i})$~M$_{\sun}$ (where $D_8$ is the distance to the source in units of 8~Mpc). If $i>30\degr$, which is likely, given the strong X-ray orbital modulation, the above relation implies $M>3$~M$_{\sun}$. As mentioned above, if at maximum the source is also radiating at a fraction ($\la$$0.5$) of the Eddington luminosity, this would lead to a BH mass $M>5$--10~M$_{\sun}$ (for a He or C/O WR star). Therefore, the compact object is consistently inferred to be a BH, although the presently available data do not allow us to further constrain its mass. However, having detected a modulation interpreted as the orbital period of the system, \src\ becomes a good candidate for searching the optical counterpart and for attempting direct dynamical measurements of the BH mass (e.g. IC 10 X--1; \citealt{prestwich07,silverman08}).

Transient ULXs and objects like \src\ show that the populations of bright X-ray binaries and low-luminosity (a few $10^{39}$~\lum) ULXs may be smoothly connected, with borderline objects and sources transiting from one class to the other depending on the varying accretion rate. However, only sources with relatively massive BHs and/or crossing the Eddington limit may become bright (isotropic $L_{\mathrm{X}}\ga10^{40}$~\lum) ULXs and show the typical spectral signatures related to the ultraluminous regime (\citealt{grd09}; Pintore et al., submitted).

Bright HMXBs and ULXs appear to be preferentially associated to low metallicity, star forming environments (e.g. \citealt{grimm03,mapelli10,mrzc11}). As discussed in the next section, NGC 4490 has both a low metallicity and a high star formation rate (SFR). Besides the transitional object \src, it contains other eight sources catalogued as ULXs \citep{roberts02,winter06,fridriksson08,gladstone09,yoshida10}. The detected number of sources and the HMXB nature of \src\ are consistent with theoretical expectations based on the properties of the environment.

\subsection{\src\ and its environment}

NGC\,4490 is a star forming, interacting galaxy. An estimated average SFR of 5.5~M$_{\sun}$ yr$^{-1}$ can be obtained from H$\alpha$ (assuming the calibration by \citealt{kennicutt87}, and an extinction of $\approx$1 mag; \citealt*{clemens99}). The companion galaxy, NGC\,4485, is connected with NGC\,4490 by an evident optical bridge. Previous studies \citep{roberts02,fridriksson08,gladstone09,yoshida10} found eight ULXs in the NGC 4485/4490 complex. \src\ probably was not included in this sample of ULXs because its luminosity is borderline and variable. 

\citet[][and references therein]{smith12} suggest an enhancement (by a factor of 2--4) of the number of ULXs in interacting galaxies. On the other hand, NGC\,4485/4490  nicely follows the correlation between SFR and ULX number (\citealt{mapelli10}, see also \citealt{grimm03,ranalli03,mineo12}):  for an SFR of $\approx$5.5 M$_{\sun}$ yr$^{-1}$, we expect 6$^{+6}_{-3}$ ULXs, fairly consistent with the observed number of eight or nine (with \src) ULXs. 

\src\ is close to the tail that connects NGC\,4490 with the smaller companion NGC\,4485, but is neither in a star-forming region nor close to a young star cluster (as it appears from the Sloan Digital Sky Survey Data Release 7  data base; \citealt{abazajian09}). A large fraction of HMXBs and ULXs were found to be offset with respect to the closest star forming region and/or young star cluster (e.g. \citealt{zezas02,kaaret04,berghea09,poutanen13}). The most likely explanation is that a number of X-ray binaries were ejected from the parent star cluster as a consequence of either supernova kick or dynamical interaction (e.g. \citealt{gualandris05,fragos09,mapelli11}; \citealt*{repetto12}).

Finally, NGC\,4490 has relatively low metallicity \mbox{($12+\log{\rm (O/H)}\approx8.3$--8.4}, corresponding to $Z\approx0.25~Z_{\sun}$, assuming  $Z_{\sun}=0.02$; \citealt{pilyugin07}). Recent studies (e.g. \citealt*{mapelli09}; \citealt{mapelli10,mrzc11,kaaret13,prestwich13}) indicate that ULXs and bright HMXBs tend to prefer metal-poor environments. The low metallicity of NGC\,4490 fairly confirms this behaviour (NGC\,4485/4490 was already included in the galaxy sample by \citealt{mapelli10}). The anti-correlation between ULXs and metallicity has been attributed either to the fact that more HMXBs can form at low-metallicity (e.g. \citealt{linden10}), or that the mass of BHs can be higher in metal-poor environments (e.g. \citealt{zampieri09,belczynski10,mapelli13}). Thus, constraining the mass of the BH candidate in CXOUJ123030.3+413853 would provide an important clue to understand the metallicity--HMXB connection.

\section*{Acknowledgements} 
This research is based on data and software provided by the \cxo\ X-Ray Center (operated for NASA by SAO) and the ESA's \xmm\ Science Operations Centre. MM and LZ acknowledge financial support from the Italian Ministry of Education, University and Research through grant FIRB 2012 (`New perspectives on the violent Universe: unveiling the physics of compact objects with joint observations of gravitational waves and electromagnetic radiation'), and  from INAF through grant \mbox{PRIN-2011-1} (`Challenging ultraluminous X-ray sources: chasing their black holes and formation pathways').

\bibliographystyle{mn2e}
\bibliography{biblio}

\bsp

\label{lastpage}

\end{document}